# A study on Deep Convolutional Neural Networks, transfer learning, and Mnet model for Cervical Cancer Detection

**Saifuddin Sagor**
Department of Computer Science and Engineering
Daffodil International University, Dhaka, Bangladesh
sagor15-6398@s.diu.edu.bd

**Dr. Md Taimur Ahad**
School of Mathematics, Physics and Computing
Toowoomba Campus
University of Southern Queensland
MdTaimur.Ahad@unisq.edu.au

**Faruk Ahmed**
Department of Computer Science and Engineering
Daffodil International University, Dhaka, Bangladesh
faruk15-4205@diu.edu.bd

**Rokonozzaman Ayon**
Department of Computer Science and Engineering
Daffodil International University, Dhaka, Bangladesh
ayon15-4393@diu.edu.bd

**Sanzida Parvin**
Department of Computer Science and Engineering
Daffodil International University, Dhaka, Bangladesh
parvin15-6265@s.diu.edu.bd

# Abstract

Cervical cancer remains one of the most common cancers affecting women worldwide, particularly in low-resource settings. Early and accurate detection through Pap smear analysis is critical to improving patient outcomes and reducing mortality. Although state-of-the-art (SOTA) Convolutional Neural Networks (CNNs) have revolutionized disease diagnosis, most SOTA CNNs are designed for large-scale object detection and classification tasks. As a result, they require substantial computational resources, extended training time, and large datasets. In this study, a lightweight CNN model, S-Net (Simple Net), is developed specifically for cervical cancer detection and classification using Pap smear images to address these limitations. Alongside S-Net, six SOTA CNNs were evaluated using transfer learning, including multi-path (DenseNet201, ResNet152), depth-based (Serasnet152), width-based multi-connection (Xception), depth-wise separable convolutions (MobileNetV2), and spatial exploitation-based (VGG19). All models, including S-Net, achieved comparable accuracy, with S-Net reaching 99.99%. However, S-Net significantly outperforms the SOTA CNNs in terms of computational efficiency and inference time, making it a more practical choice for real-time and resource-constrained applications. A major limitation in CNN-based medical diagnosis remains the lack of transparency in the decision-making process. To address this, Explainable AI (XAI) techniques, such as SHAP, LIME, and Grad-CAM, were employed to visualize and interpret the key image regions influencing model predictions. The novelty of this study lies in the development of a highly accurate yet computationally lightweight model (S-Net) caPable of rapid inference while maintaining interpretability through XAI integration. Furthermore, this work analyzes the behavior of SOTA CNNs, investigates the effects of negative transfer learning on Pap smear images, and examines pixel intensity patterns in correctly and incorrectly classified samples.

# 1. Introduction

Cervical cancer, which affects the cervix at the lower end of the uterus, remains one of the most common cancers among women globally. The World Health Organization (WHO) reports that cervical cancer is a leading cause of cancer-related deaths in over 40 countries, highlighting its public health significance (Joynab et al., 2024; Fan et al., 2023). In 2020, around 0.6 million cases of cervical cancer were diagnosed worldwide (Sarhangi et al., 2024). However, traditional screening methods are prone to high false-positive rates due to human error, compromising the accuracy of diagnosis and early detection.

To overcome these challenges, machine learning (ML) and deep learning (DL) based computer-aided diagnostic (CAD) techniques have been increasingly utilized for the automatic analysis of cervical cytology and colposcopy images. These AI-driven technologies are significantly improving diagnostic accuracy by automating image segmentation and classification, thus reducing reliance on manual analysis and minimizing human error ( Ahad et al., 2023; Mustofa et al., 2023; Bhowmik et al., 2024, Ahmed & Ahad, 2023; Emon & Ahad, 2024; Mustofa et al., 2024; Preanto et al., 2024; Mamun et al., 2023; Ahad et al., 2024; Mustofa et al., 2025; Preanto et al., 2024; Ahmed et al., 2023; Ahad et al., 2024; Bhowmik et al., 2023; Ahad et al., 2024; Mamun et al., 2025; Ahad et al., 2024, Ahad et al., 2024; Islam et al., 2024; Ahad et al., 2024; Ahmed et al., 2024; Ahad et al., 2024; Preanto et al., 2024; Preanto et al., 2024; Ahad et al., 2024; Ahad et al., 2024; Ahad et al., 2024; Mamun et al., 2024; Emon et al., 2023; Emon et al., 2023; Biplob et al., 2023; Ahad et al., 2023; Ahad et al., 2023; Ahad et al., 2023; Ahad et al., 2023; Ahad et al., 2023; Youneszade et al., 2023). This represents a critical advancement in cervical cancer diagnosis, enhancing the effectiveness of screening and improving early detection rates.

One of the most promising AI models in cervical cancer detection is the Deep Convolutional Neural Network (DCNN). DCNNs are particularly effective in classifying early-stage cancer and detecting malignant cells by automatically identifying complex patterns in medical images (Kumar et al., 2024; Nirmala et al., 2024). The incorporation of transfer learning, using pre-trained models like those from ImageNet, further enhances performance by allowing the models to adapt to smaller, task-specific datasets (Morid et al., 2021; Atasever et al., 2023).

However, DCNNs present challenges related to their high computational demands and large memory footprint, which can limit their application, especially in low-resource settings (Pacal et al., 2024; Lau et al., 2024). To address this, researchers have developed lightweight CNN models that offer comparable performance while reducing the complexity of the network. These models are more practical for deployment in mobile devices and resource-constrained environments, providing an accessible solution for real-world clinical settings (Mathivanan et al., 2024; Gendy et al., 2024; Mehedi et al., 2024; Dogani et al., 2023).

A key challenge in medical AI is the lack of interpretability of deep learning models. To improve model transparency and trust, Explainable AI (XAI) methods such as Grad-CAM, LIME, and SHAP have been developed. These methods provide insights into model decisions, enhancing clinician trust and supporting ethical decision-making in clinical practice (Arrieta et al., 2020; Albahri et al., 2023; Band et al., 2023).

Despite notable advancements in CNN models for cervical cancer detection, several knowledge gaps persist:

1. Many studies focus on developing CNN models but lack attention to clinical concerns like model generalization and interpretability (Zhang et al., 2025; Rahman et al., 2023).
2. Well-known CNN architectures have not been rigorously evaluated for cervical cancer detection in real-world clinical environments (Sambyal et al., 2023).
3. Despite its success in other domains, transfer learning remains underutilized in cervical cancer CAD systems (Yeasmin et al., 2024).
4. High false-positive rates and low reliability limit the clinical applicability of existing CNN-based methods (Attallah, 2023; Painuli & Bhardwaj, 2022).

Following the gaps, this study conducts three experiments, which make the following contributions and novelties of this study:

1. A lightweight CNN model, S-Net, is introduced for efficient cervical cancer detection from Pap smear images, optimizing both computational efficiency and accuracy. Additionally, six state-of-the-art CNN architectures (VGG19, ResNet152v2, SE-ResNet152, MobileNetV2, Xception, DenseNet201) were evaluated to identify the most effective model for this task
2. The study applies XAI techniques (LIME, SHAP, Grad-CAM) to improve model interpretability and transparency, while statistical analysis of pixel intensity evaluates classification outcomes (TP, FP, TN, FN).

3. pixel intensity as a key factor in image classification, investigating its role in cervical cancer detection and its impact on classification accuracy.

## 2. Related works

Tan et al. (2024) and Khowaja et al. (2024) focused on using deep learning models for cervical cancer detection through Pap smear images. Tan et al. (2024) employed transfer learning with pre-trained CNN models for a seven-class classification task, achieving the best performance with DenseNet-201. Similarly, Khowaja et al. developed a framework that outperformed 25 other models, showing impressive accuracy on the Mendeley and SIPaKMeD datasets. Their work emphasizes the potential of these models to improve cervical cancer screening and reduce misdiagnosis.

Deo et al. (2024) introduced CerviFormer, a Transformer-based model, which excelled at classifying Pap smear images, particularly for large-scale inputs. This method demonstrated competitive results on two public datasets, with accuracy rates of 96.67% and 94.57%, showing promise compared to traditional CNN-based methods. Additionally, Pacal (2024) utilized 106 deep learning models, including CNNs and vision transformers, to achieve remarkable accuracy on the SIPaKMeD and LBC datasets, surpassing existing models in both datasets.

Mazumder et al. (2024) and Nour et al. (2024) combined CNNs like InceptionV3 and EfficientNetV2S to enhance cervical cancer detection accuracy. These models demonstrated accuracy rates of up to 99.98%, outperforming contemporary approaches. Hybrid methods, combining deep learning with handcrafted features, were also explored by Joseph et al. (2024) for early-stage cancer detection, yielding a high median recall of 99.5%.

Fan et al. (2023) and Tomko et al. (2022) worked on optimizing models for better performance. Fan et al. introduced CAM-VT, a weakly supervised model using conjugated attention and visual transformers, which improved pap slide identification accuracy. Tomko et al.(2022) focused on optimizing input image sizes for models like EfficientNetB0, improving classification performance. Additionally, some studies, like those of Prasanthi Shandilya et al. (2024), employed lightweight CNNs (e.g., MobileNetV2 and ResNet-18) to achieve high accuracy while reducing computational complexity, making them more suitable for clinical applications.

Wubineh et al. (2024) and Hussain et al. (2024) explored segmentation techniques and model interpretability. Wubineh et al. focused on cytoplasm and nuclei segmentation for cervical cancer detection, with EfficientNetB2 achieving impressive accuracy. Meanwhile, Hussain et al.(2024) used the Xception model for classifying cervical cancer, with AUC scores of up to 0.98. These studies highlight the importance of model transparency and accurate feature localization, improving the trustworthiness and effectiveness of AI-based cervical cancer diagnostic tools.

Compared to the D-CNN and transfer learning models, the M-Net model is superior in accuracy. Both D-CNN and transfer learning methods are computationally expensive and heavy. The researcher put forth an M-Net model in this investigation. Shorter models, less convolution, fewer fully linked layers, and fewer layers comprise the M-Net model. This M-Net approach reduced study time waste while improving accuracy for cervical cancer. Realizing the effectiveness of CNN in the detection and classification of cervical cancer, scholars such as Leung and Yue (2024), Luo et al. (2021), Azad et al. (2024), Uyanga et al. (2024), and Huang et al. (2025) have conducted extensive research in this area. They experimented with CNN and its variants to improve cervical cancer detection and classification.

# 3. Experimental method and materials

This section describes the hardware specification, dataset description, S-Net model development, and training procedure for this study.

This study presents a comprehensive cervical cancer detection pipeline using Pap smear images. Initially, the images undergo preprocessing before classification. The framework evaluates six state-of-the-art CNN models, VGG19, ResNet152V2, SE-ResNet152, DenseNet201, MobileNetV2, and Xception, through two experimental setups: (i) training from scratch and (ii) transfer learning using ImageNet weights. Additionally, a custom lightweight CNN model, S-Net, is introduced and assessed through 5-fold cross-validation and explainable AI (XAI) techniques like LIME, SHAP, and Grad-CAM.

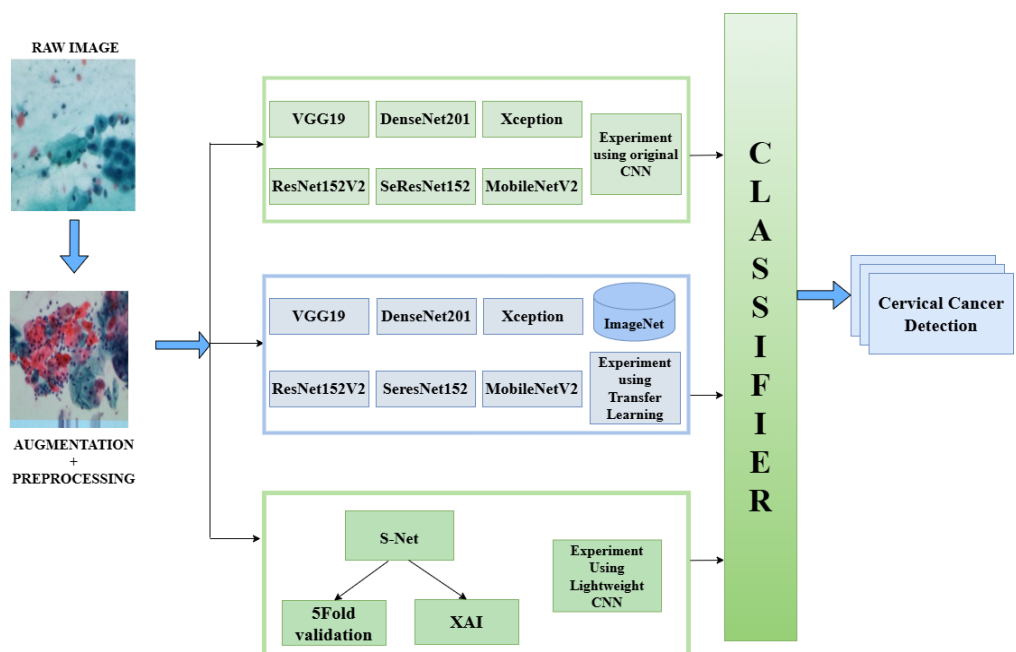

Figure 1: Experimental flow of the study.

## 3.1 Hardware Specification

The experiments were conducted using Precision 7680 Workstation. The workstation is a 13th-generation Intel® Core™ i9-13950HX vPro with a Windows 11 Pro operating system and NVIDIA® RTX™ 3500 Ada Generation, 32 GB DDR5, and 1 TB SSD. Python (version 3.9) was chosen as the programming language as the version supported the TensorFlow-GPU, SHAP, and LIME generation.

## 3.2 Dataset description

The dataset for the study was collected from a public repository and the dataset is Obuli Sai Naren. (2022) Multi Cancer Dataset. Figure 1 displays samples of images used in the study. This dataset contains 25000 Pap smear (Papanicolaou smear) microscopic images and is classified into five (5) classes: Cervix_Dyskeratotic (Dyk), Cervix_Koilocytotic (Koc), Cervix_Metaplastic (Mep), Cervix_Parabasal (Pab) and Cervix_Superficial Moderate (Sfi). Cervical Cancer images were taken from the Obuli Sai Naren. (2022). The images were captured using pap smears and stored in JPG format.

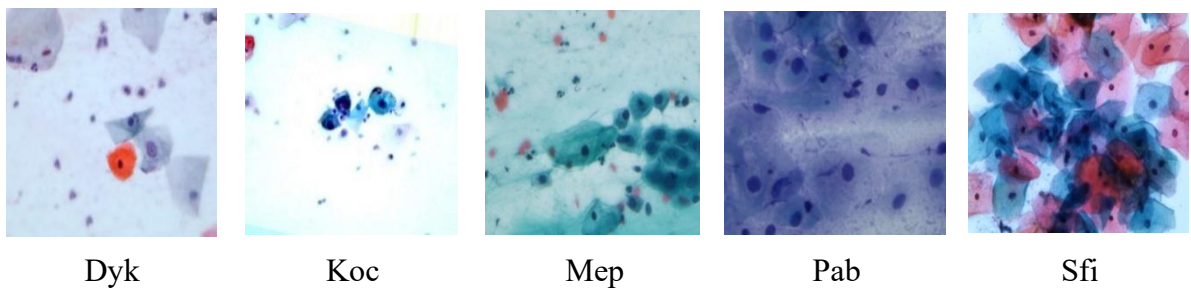

| Dyk | Koc | Mep | Pab | Sfi |

Figure 2: Example of 5 cervical cancer classes

## 3.3 Image Augmentation

This study used position augmentation, such as scaling, cropping, flipping, and rotation, and color augmentation, such as brightness, contrast, and saturation, was deployed. Random rotation from −15 to 15 degrees, rotations of multiples of 90 degrees at random, random distortion, shear transformation, vertical flip, horizontal flip, skewing, and intensity transformation were also used in the data augmentation process.

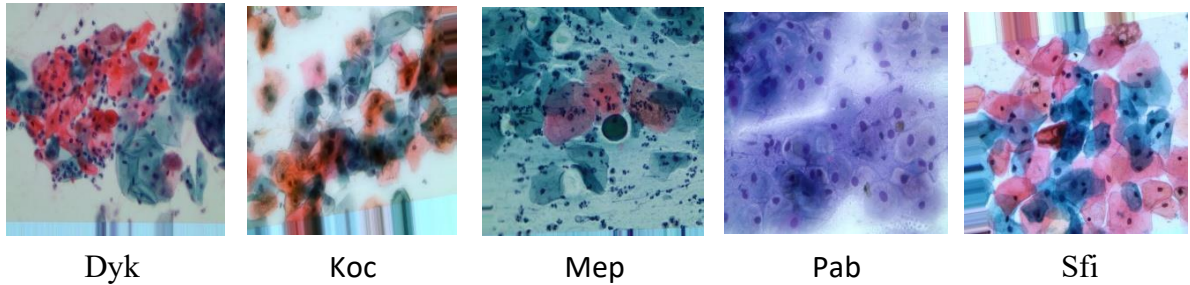

Dyk　　　　　Koc　　　　　Mep　　　　　Pab　　　　　Sfi

Figure 3. Image after augmentation

## 3.4 Model Development

Inspired by the architectural taxonomy of CNNs proposed by Khan et al. (2020), we selected six diverse and high-performing convolutional neural networks (CNNs): VGG19, ResNet152, SE-ResNet152, DenseNet201, Xception, and MobileNetV2 to evaluate their effectiveness in classifying cervical cancer cytology images. Each model reflects a unique architectural class: spatial exploitation (VGG19), depth-based learning (ResNet152), attention mechanisms (SE-ResNet152), multi-path connectivity (DenseNet201), width-based design (Xception), and efficient lightweight architecture (MobileNetV2).

To address the challenge of limited annotated medical data, we employed transfer learning, fine-tuning each model pre-trained on ImageNet. The experimental results demonstrate that advanced architectures, particularly those integrating attention and residual connections such as SE-ResNet152 (99.81%) and MobileNetV2 (99.53%), significantly outperform older architectures like VGG19 (88.58%). These findings confirm that transfer learning, when combined with modern CNN designs, substantially improves diagnostic accuracy in cytological image analysis.

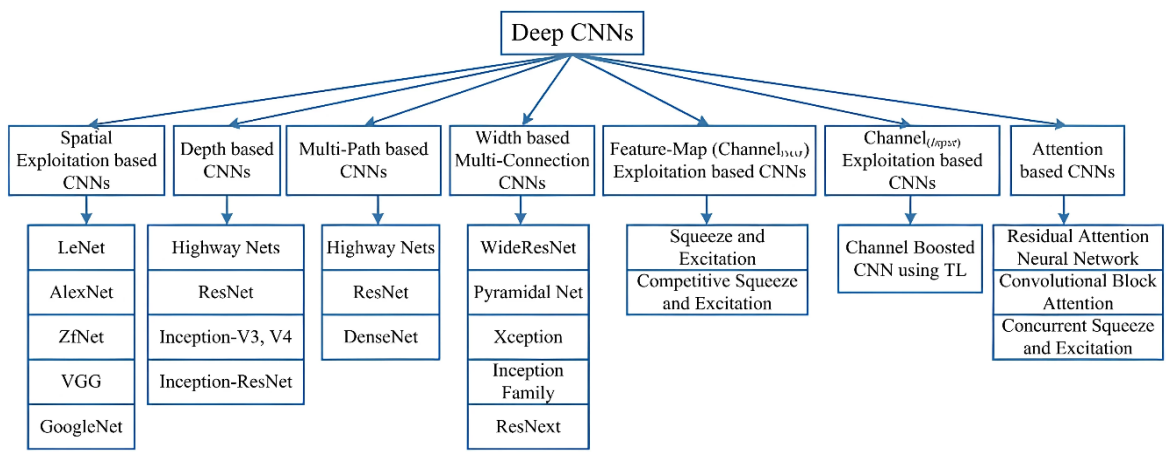

Figure 4: Classification of Deep CNN Architectures.

## 4. Experimental description and results

This study conducted three experiments to detect and classify cervical cancer pap smear images. Firstly, a lightweight CNN S-Net; secondly, six SOTA CNN's Multipath (DenseNet201), Depth-based (ResNet152), width-based multi-connection (Xception), depth-wise separable convolutions (MobileNetV2), spatial exploitation based (VGG19); thirdly, transfer learning of 6 SOTA CNN's a Multi path-depth-Width based (DenseNet201- Xception) and Multi path-depth-Spatial exploitation (ResNet152-VGG19). The experimental processes and results are described below.

### 4.1 Experiment 1: S-Net development process and results

The S-Net model, training process, and results and discussed below:

#### 4.1.1 S-Net model development

The S-Net model, designed for cervical cancer detection in Pap smear images, begins with a 2D convolutional layer featuring 48 filters, followed by max-pooling to reduce spatial dimensions. The architecture deepens progressively with convolutional layers containing 128, 192, and 512 filters, interspersed with pooling layers for hierarchical feature extraction. A fully connected layer with 1024 neurons is employed, followed by dropout regularisation, culminating in a dense output layer with five neurons for multi-class classification (Dyk, Koc, Mep, Pab, Sfi). With a total of 2,020,549 trainable parameters, the model is optimised for high-capacity learning and multi-class image categorisation.

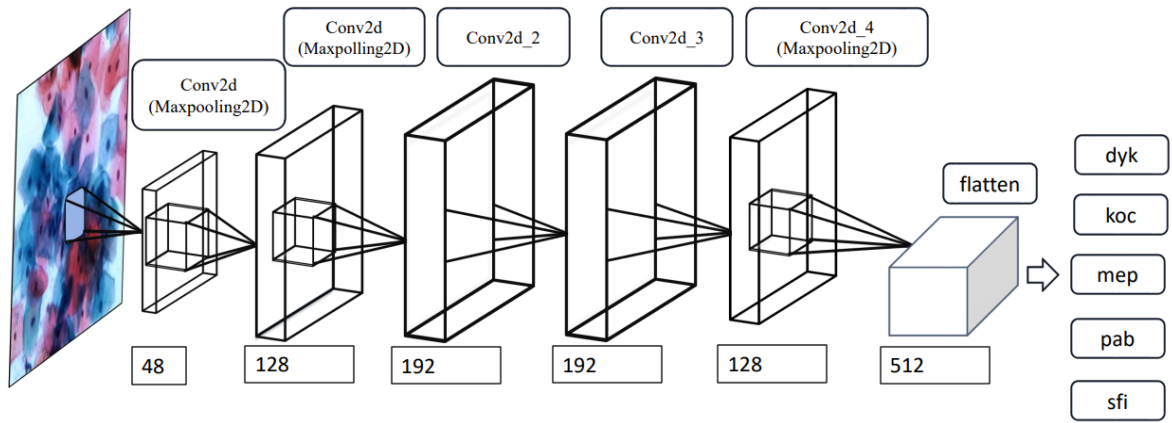

Figure 5: S-Net model visualisation

### 4.1.2 Model Summary

Table 1: Hyperparameters

| Layer | Output Shape | Parameters | Description |
|---|---|---|---|
| Conv2D | (None, 52, 52, 48) | 3,648 | Convolutional layer with 48 filters, kernel size (3,3) |
| MaxPooling2D | (None, 17, 17, 48) | 0 | Max pooling operation (2x2) |
| Conv2D | (None, 17, 17, 128) | 55,424 | Convolutional layer with 128 filters, kernel size (3,3) |
| MaxPooling2D | (None, 5, 5, 128) | 0 | Max pooling operation (2x2) |
| Conv2D | (None, 5, 5, 192) | 221,376 | Convolutional layer with 192 filters, kernel size (3,3) |
| Conv2D | (None, 5, 5, 192) | 331,968 | Convolutional layer with 192 filters, kernel size (3,3) |
| Conv2D | (None, 5, 5, 128) | 221,312 | Convolutional layer with 128 filters, kernel size (3,3) |
| MaxPooling2D | (None, 1, 1, 128) | 0 | Max pooling operation (5x5) |
| Flatten | (None, 128) | 0 | Flatten the output for input into dense layers |
| Dense | (None, 1024) | 132,096 | Fully connected layer with 1024 units |
| Dropout | (None, 1024) | 0 | Dropout for regularization (no parameters) |
| Dense | (None, 1024) | 1,049,600 | Fully connected layer with 1024 units |
| Dense | (None, 5) | 5,125 | Output layer with 5 units (classification output) |
| Total Parameters | | 2,020,549 | Total number of trainable parameters |
| Trainable Parameters | | 2,020,549 | All parameters in the model are trainable |
| Non-trainable Parameters | | 0 | No non-trainable parameters |

### 4.1.3 Training Setup

The S-Net model was trained using K-fold cross-validation, with K set to 5. Each fold underwent 50 training epochs, with a fixed batch size of 16. The Adam optimization algorithm was used for model optimization. The algorithm ensured the adaptability of the model weights based on the calculated gradients, thereby enhancing performance and accuracy in classifying cervical cancer. The loss function was sparse categorical cross-entropy. Early stopping and a learning rate scheduler were applied during training. The training hyperparameters are provided in Table 1.

Table 1: Hyperparameters of training

| Parameters |
|---|
| Epochs = 50 |
| Batch size = 16 |
| Image size = (64, 64, 3) |
| Learning rate = 1.0000e-04 |
| K_folds = 5 |
| Optimizer = Adam(learning_rate=LEARNING_RATE) |
| Loss = SparseCategoricalCrossentropy(from_logits=True) |
| Early stopping = EarlyStopping(monitor='val_accuracy', patience=10, verbose=1, restore_best_weights=True) |
| Learning_rate_scheduler = LearningRateScheduler(lambda epoch: LEARNING_RATE * 0.1 ** (epoch // 10)) |
| Callbacks = [early_stopping, lr_scheduler] |

### 4.1.4 Results of the S-Net

The S-Net model demonstrated high classification accuracy across all metrics (precision, recall, F1-score, and support). In Fold 1, minor misclassifications occurred with "Mep" and "Koc" samples, but overall accuracy was strong. Fold 2 showed improved precision, with no misclassification in "Koc," "Pab," or "Sfi" classes. Folds 3 to 5 exhibited near-perfect classification, minimizing errors. The confusion matrices confirm the model's robustness and suitability for clinical diagnostics.

Table 2: Fold-Wise Classification report with epochs of S-Net

| Fold | Class | Precision | Recall | F1-Score | Support |
|---|---|---|---|---|---|
| 1 | cervix_Dyk | 1.00 | 1.00 | 1.00 | 1548 |
| | cervix_Koc | 1.00 | 1.00 | 1.00 | 1513 |
| | cervix_Mep | 1.00 | 1.00 | 1.00 | 1504 |
| | cervix_Pab | 1.00 | 1.00 | 1.00 | 1594 |

|   |              | precision | recall | f1-score | support |
|---|--------------|-----------|--------|----------|---------|
|   | cervix_Sfi   | 1.00      | 1.00   | 1.00     | 1525    |
|   | Accuracy     |           |        | 1.00     | 7684    |
|   | Macro Avg    | 1.00      | 1.00   | 1.00     | 7684    |
|   | Weighted Avg | 1.00      | 1.00   | 1.00     | 7684    |
| 2 | cervix_Dyk   | 1.00      | 1.00   | 1.00     | 1530    |
|   | cervix_Koc   | 1.00      | 1.00   | 1.00     | 1599    |
|   | cervix_Mep   | 1.00      | 1.00   | 1.00     | 1555    |
|   | cervix_Pab   | 1.00      | 1.00   | 1.00     | 1481    |
|   | cervix_Sfi   | 1.00      | 1.00   | 1.00     | 1518    |
|   | Accuracy     |           |        | 1.00     | 7683    |
|   | Macro Avg    | 1.00      | 1.00   | 1.00     | 7683    |
|   | Weighted Avg | 1.00      | 1.00   | 1.00     | 7683    |
| 3 | cervix_Dyk   | 1.00      | 1.00   | 1.00     | 1541    |
|   | cervix_Koc   | 1.00      | 1.00   | 1.00     | 1536    |
|   | cervix_Mep   | 1.00      | 1.00   | 1.00     | 1496    |
|   | cervix_Pab   | 1.00      | 1.00   | 1.00     | 1550    |
|   | cervix_Sfi   | 1.00      | 1.00   | 1.00     | 1560    |
|   | Accuracy     |           |        | 1.00     | 7683    |
|   | Macro Avg    | 1.00      | 1.00   | 1.00     | 7683    |
|   | Weighted Avg | 1.00      | 1.00   | 1.00     | 7683    |
| 4 | cervix_Dyk   | 1.00      | 1.00   | 1.00     | 1549    |
|   | cervix_Koc   | 1.00      | 1.00   | 1.00     | 1491    |
|   | cervix_Mep   | 1.00      | 1.00   | 1.00     | 1571    |
|   | cervix_Pab   | 1.00      | 1.00   | 1.00     | 1551    |
|   | cervix_Sfi   | 1.00      | 1.00   | 1.00     | 1521    |
|   | Accuracy     |           |        | 1.00     | 7683    |
|   | Macro Avg    | 1.00      | 1.00   | 1.00     | 7683    |
|   | Weighted Avg | 1.00      | 1.00   | 1.00     | 7683    |
|   | cervix_Dyk   | 0.99      | 0.99   | 0.99     | 1810    |
| 5 | cervix_Dyk   | 1.00      | 1.00   | 1.00     | 1562    |
|   | cervix_Koc   | 1.00      | 1.00   | 1.00     | 1525    |
|   | cervix_Mep   | 1.00      | 1.00   | 1.00     | 1531    |
|   | cervix_Pab   | 1.00      | 1.00   | 1.00     | 1520    |
|   | cervix_Sfi   | 1.00      | 1.00   | 1.00     | 1545    |
|   | Accuracy     |           |        | 1.00     | 7683    |
|   | Macro Avg    | 1.00      | 1.00   | 1.00     | 7683    |
|   | Weighted Avg | 1.00      | 1.00   | 1.00     | 7683    |

Dyskeratotic=(Dyk) Koilocytotic=(Koc) Metaplastic= (Mep) Parabasal=(Pab) Superficial Moderate=(Sfi).

**Confusion Matrix Fold 1**

|     | Dyk  | Koc  | Mep  | Pab  | Sfi  |
|-----|------|------|------|------|------|
| Dyk | 1548 | 0    | 0    | 0    | 0    |
| Koc | 2    | 1511 | 0    | 0    | 0    |
| Mep | 2    | 0    | 1502 | 0    | 0    |
| Pab | 0    | 0    | 0    | 1594 | 0    |
| Sfi | 0    | 0    | 0    | 0    | 1525 |

**Confusion Matrix Fold 2**

|     | Dyk  | Koc  | Mep  | Pab  | Sfi  |
|-----|------|------|------|------|------|
| Dyk | 1530 | 0    | 0    | 0    | 0    |
| Koc | 0    | 1599 | 0    | 0    | 0    |
| Mep | 0    | 0    | 1555 | 0    | 0    |
| Pab | 0    | 0    | 0    | 1581 | 0    |
| Sfi | 0    | 0    | 0    | 0    | 1518 |

**Confusion Matrix Fold 3**

|     | Dyk  | Koc  | Mep  | Pab  | Sfi  |
|-----|------|------|------|------|------|
| Dyk | 1541 | 0    | 0    | 0    | 0    |
| Koc | 0    | 1536 | 0    | 0    | 0    |
| Mep | 0    | 0    | 1496 | 0    | 0    |
| Pab | 0    | 0    | 0    | 1550 | 0    |
| Sfi | 0    | 0    | 0    | 0    | 1560 |

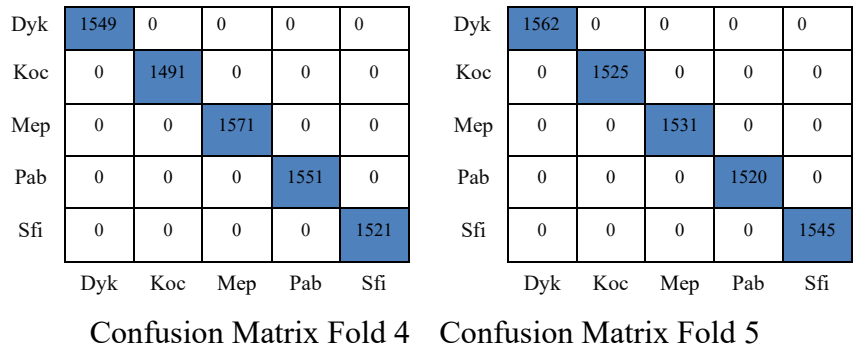

Confusion Matrix Fold 4    Confusion Matrix Fold 5

Figure 6: Fold-wise confusion matrix of S-Net

Figure 6 Confusion matrices from five-fold cross-validation reveal the strong classification ability of the S-Net model across five cervical cell types. Each fold shows a clear diagonal dominance, indicating high prediction accuracy. Minor misclassifications are observed in Fold 1, primarily between morphologically similar classes. Folds 2 to 5 exhibit near-perfect classification performance. These results highlight S-Net's robustness and consistency in multi-class cervical cell classification.

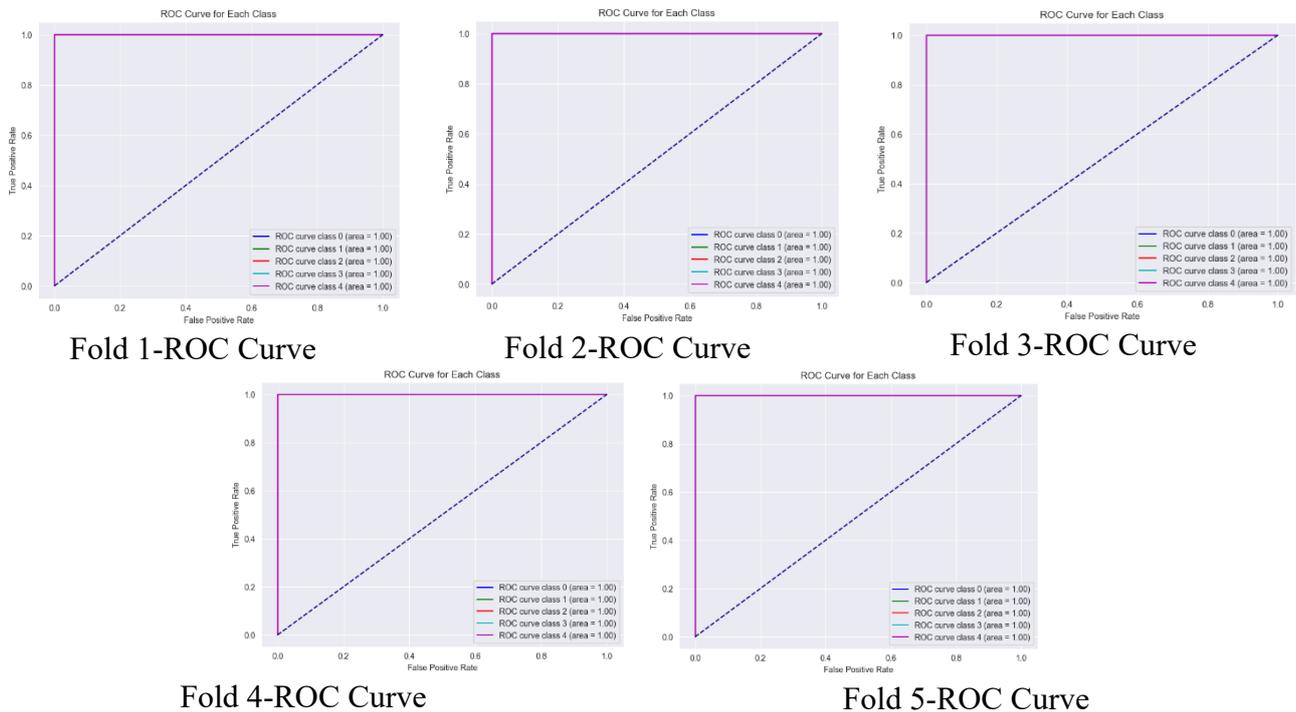

Fold 1-ROC Curve    Fold 2-ROC Curve    Fold 3-ROC Curve

Fold 4-ROC Curve    Fold 5-ROC Curve

Figure 7: Fold-wise confusion matrix with Epochs and Accuracy

Figure 7 ROC curves from five-fold cross-validation demonstrate S-Net's strong classification performance. In Fold 1, the model effectively minimizes false positives while maintaining high true positive rates. Fold 2 shows improved precision with tighter ROC curves. Folds 3 to 5

confirm the model's consistent and robust classification accuracy. Overall, these results emphasize the reliability of S-Net in cervical cancer detection.

|     | Dyk  | Koc  | Mep  | Pab  | Sfi  |
|-----|------|------|------|------|------|
| Dyk | 7730 | 0    | 0    | 0    | 0    |
| Koc | 2    | 7662 | 0    | 0    | 0    |
| Mep | 2    | 0    | 7655 | 0    | 0    |
| Pab | 0    | 0    | 0    | 7696 | 0    |
| Sfi | 0    | 0    | 0    | 0    | 7669 |

| Class        | Precision | Recall | F1-Score | Support |
|--------------|-----------|--------|----------|---------|
| cervix_Dyk   | 1.00      | 1.00   | 1.00     | 7730    |
| cervix_Koc   | 1.00      | 1.00   | 1.00     | 7664    |
| cervix_Mep   | 1.00      | 1.00   | 1.00     | 7657    |
| cervix_Pab   | 1.00      | 1.00   | 1.00     | 7696    |
| cervix_Sfi   | 1.00      | 1.00   | 1.00     | 7669    |
| Accuracy     |           |        | 1.00     | 38416   |
| Macro Avg    | 1.00      | 1.00   | 1.00     | 38416   |
| Weighted Avg | 1.00      | 1.00   | 1.00     | 38416   |

Figure 8: Combined confusion matrix of M-Net

Figure 8 The combined classification metrics for the S-Net model demonstrate exceptional performance, with an overall accuracy of 99.98%. Precision, recall, and F1-scores range from 99.97% to 99.99%. Class-specific results indicate strong performance, with only minor misclassifications in certain folds.

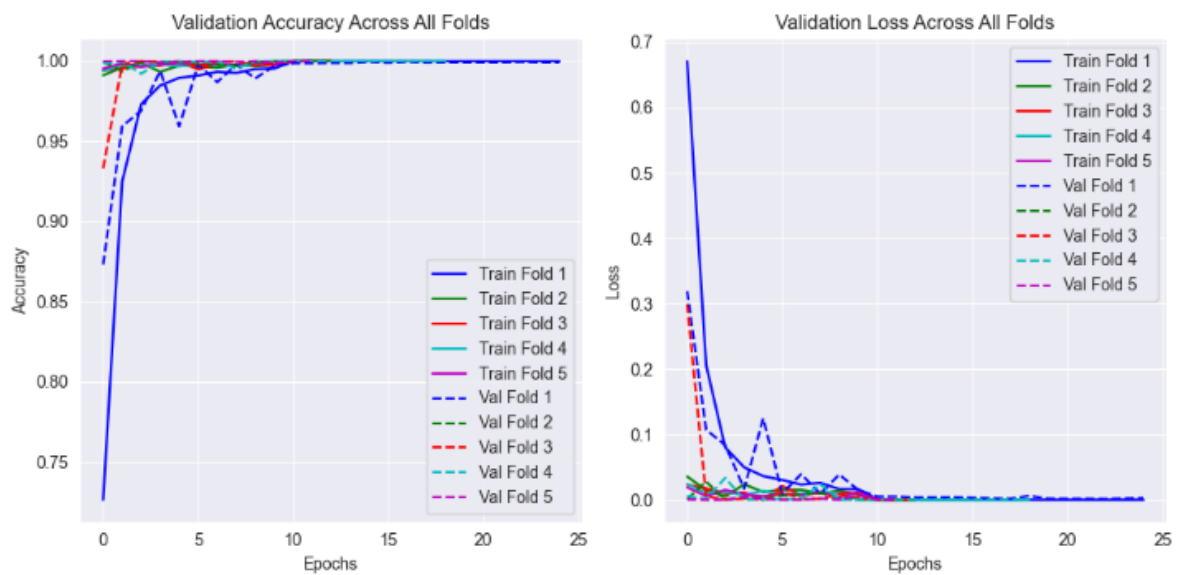

Figure 9: Fold-Wise Classification Report with Epochs and Accuracy

Figure 9 The S-Net model shows minimal classification errors and high discriminatory power, with low false positives. Accuracy curves rise steadily, and loss curves decline consistently, confirming the model's robustness, efficiency, and reliability for cervical cancer classification.

## 4.2 Experiment 2- Performance of the Original CNN Network

The performances of six original CNN architectures VGG19, ResNet152v2, SE-ResNet152, DenseNet201, Xception, and MobileNetV2, are presented in this section. Among all models, the highest accuracy was achieved, and the overall performance was excellent.

Table 3: Accuracy for classification of individual CNN networks in detecting cervical cancer. The table reports the Precision, Recall, F1-Score, and Support (n) for each network, based on the number of images (n = numbers).

|  |  | Dyk | Koc | Mep | Pab | Sfi | Model Accuracy |
|---|---|---|---|---|---|---|---|
| VGG19 | Precision | 100% | 100% | 100% | 100% | 100% | 99.97% |
|  | Recall | 100% | 100% | 100% | 100% | 100% |  |
|  | F1-score | 100% | 100% | 100% | 100% | 100% |  |
|  | Support (N) | 1485 | 1477 | 1491 | 1487 | 1484 |  |
| ResNet152 | Precision | 100% | 100% | 100% | 100% | 100% | 99.99% |
|  | Recall | 100% | 100% | 100% | 100% | 100% |  |
|  | F1-score | 100% | 100% | 100% | 100% | 100% |  |
|  | Support (N) | 1484 | 1481 | 1489 | 1480 | 1490 |  |
| Seresnet152 | Precision | 100% | 100% | 100% | 100% | 100% | 99.99% |
|  | Recall | 100% | 100% | 100% | 100% | 100% |  |
|  | F1-score | 100% | 100% | 100% | 100% | 100% |  |
|  | Support (N) | 1488 | 1491 | 1481 | 1482 | 1483 |  |
| Densenet201 | Precision | 100% | 100% | 100% | 100% | 100% | 100% |
|  | Recall | 100% | 100% | 100% | 100% | 100% |  |
|  | F1-score | 100% | 100% | 100% | 100% | 100% |  |
|  | Support (N) | 1485 | 1484 | 1485 | 1488 | 1482 |  |
| Xception | Precision | 100% | 100% | 100% | 100% | 100% | 100% |
|  | Recall | 100% | 100% | 100% | 100% | 100% |  |
|  | F1-score | 100% | 100% | 100% | 100% | 100% |  |
|  | Support (N) | 1488 | 1483 | 1481 | 1485 | 1487 |  |
| MobilenetV2 | Precision | 100% | 100% | 100% | 100% | 100% | 100% |
|  | Recall | 100% | 100% | 100% | 100% | 100% |  |
|  | F1-score | 100% | 100% | 100% | 100% | 100% |  |
|  | Support (N) | 1485 | 1482 | 1488 | 1486 | 1483 |  |

Table 3 highlights the precision values for each architecture evaluated on the test dataset. The VGG19, ResNet152v2, DenseNet201, SE-ResNet152, Xception, and MobileNetV2 architectures delivered outstanding results.

From the table, it is evident that the DenseNet201, Xception, and MobileNetV2 models excelled in correctly detecting and classifying cervical cancer compared to the other architectures. These models provided highly accurate classification and demonstrated exceptional performance in detecting cervical cancer.

Dyskeratotic=(Dyk) Koilocytotic=(Koc) Metaplastic= (Mep) Parabasal=(Pab) Superficial Moderate=(Sfi).

**VGG19 confusion matrix:**

|  | Dyk | Koc | Mep | Pab | Sfi |
|---|---|---|---|---|---|
| Dyk | 1485 | 0 | 0 | 0 | 0 |
| Koc | 0 | 1475 | 0 | 0 | 0 |
| Mep | 0 | 0 | 1491 | 0 | 0 |
| Pab | 0 | 0 | 0 | 1487 | 0 |
| Sfi | 0 | 2 | 0 | 0 | 1484 |

**DenseNet201 confusion matrix:**

|  | Dyk | Koc | Mep | Pab | Sfi |
|---|---|---|---|---|---|
| Dyk | 1485 | 0 | 0 | 0 | 0 |
| Koc | 0 | 1484 | 0 | 0 | 0 |
| Mep | 0 | 0 | 1485 | 0 | 0 |
| Pab | 0 | 0 | 0 | 1488 | 0 |
| Sfi | 0 | 0 | 0 | 0 | 1482 |

**ResNet152v2 confusion matrix:**

|  | Dyk | Koc | Mep | Pab | Sfi |
|---|---|---|---|---|---|
| Dyk | 1484 | 0 | 0 | 0 | 0 |
| Koc | 0 | 1481 | 1 | 0 | 0 |
| Mep | 0 | 0 | 1488 | 0 | 0 |
| Pab | 0 | 0 | 0 | 1480 | 0 |
| Sfi | 0 | 0 | 0 | 0 | 1490 |

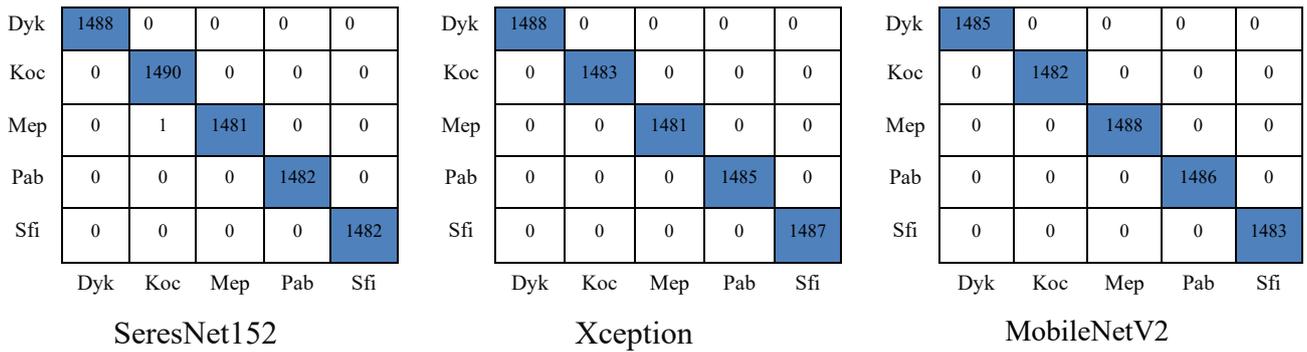

Figure 10: Six confusion matrices of original CNNs

Figure 10 shows confusion matrices for six CNNs: VGG19, DenseNet201, ResNet152v2, SEResNet152, Xception, and MobileNetV2. DenseNet201, Xception, and MobileNetV2 achieved better classification accuracy, correctly classifying 919 and 977 images, as noted in Table 3.

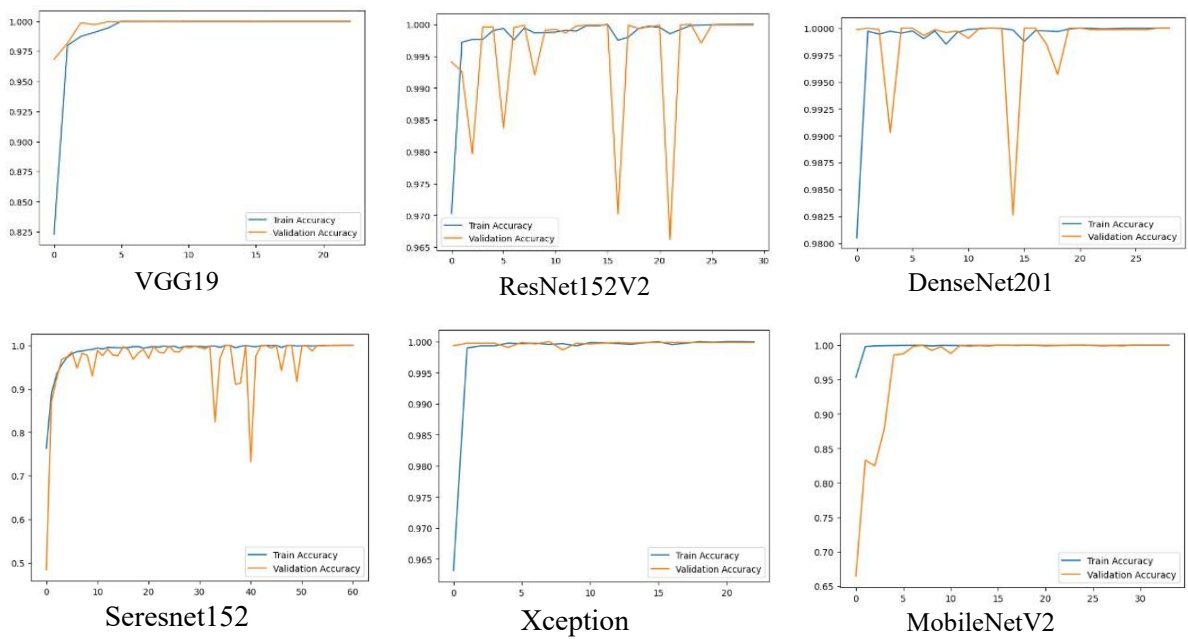

Figure 11: The training and validation accuracy of the original CNNs.

Figure 11 presents the training and validation accuracy curves for six CNNs: VGG19, ResNet152V2, DenseNet201, SE-ResNet152, Xception, and MobileNetV2. DenseNet201, Xception, and MobileNetV2 show stable accuracy with minimal fluctuations, while VGG19 converges quickly and maintains high accuracy. ResNet152V2 and SE-ResNet152 exhibit occasional drops in validation accuracy but remain strong overall.

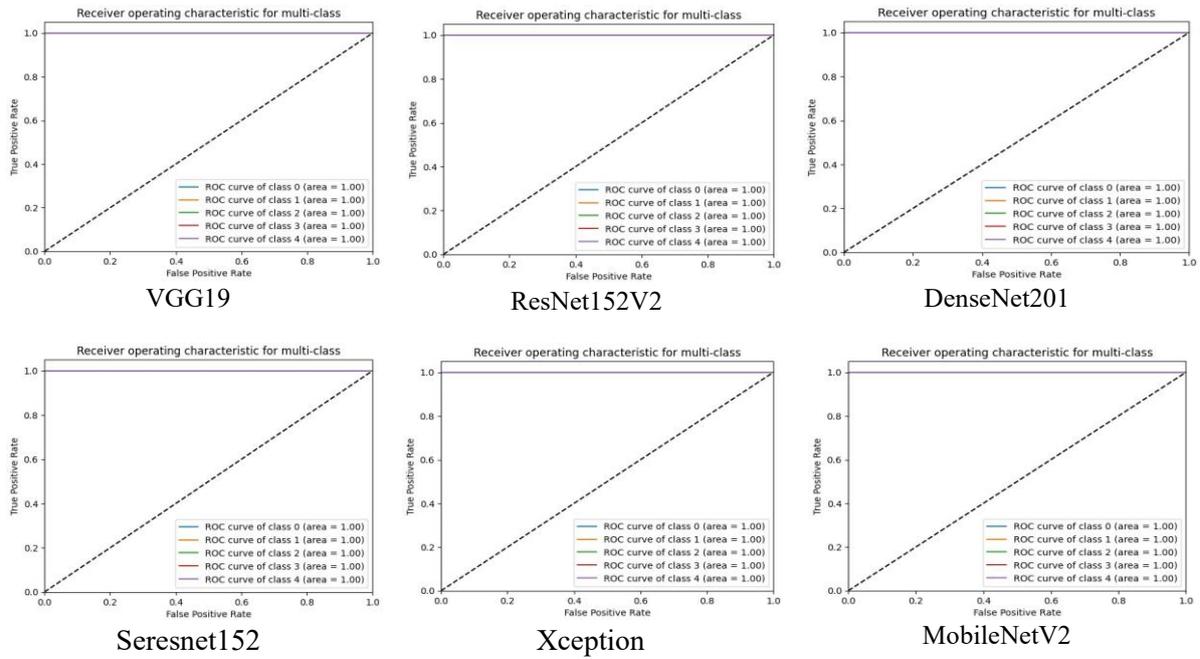

Figure 12: Six roc curves of original CNNs.

Figure 12 The ROC curves for various CNN architectures (VGG19, ResNet152V2, DenseNet201, SE-ResNet152, Xception, and MobileNetV2) show an AUC of 1.00 for all models, indicating perfect classification performance. Each model effectively distinguishes between classes with high confidence.

## 4.3 Experiment 4: Transfer Learning CNN Network Accuracy in Detecting Cervical Cancer

Table 4. Transfer learning CNN network accuracy in detecting cervical cancer.

|  |  | **Dyk** | **Koc** | **Mep** | **Pab** | **Sfi** | **Model Accuracy** |
|---|---|---|---|---|---|---|---|
| VGG19 | Precision | 87% | 85% | 83% | 94% | 94% | 88.58% |
|  | Recall | 82% | 80% | 88% | 99% | 94% |  |
|  | F1-score | 84% | 82% | 86% | 97% | 94% |  |
|  | Support (N) | 1484 | 1480 | 1487 | 1488 | 1485 |  |
|  | Precision | 99% | 99% | 99% | 100% | 99% |  |

| Model | Metric | Dyk | Koc | Mep | Pab | Sfi | Accuracy |
|---|---|---|---|---|---|---|---|
| ResNet152 | Recall | 99% | 98% | 99% | 100% | 100% | 99.16% |
| | F1-score | 99% | 98% | 99% | 100% | 99% | |
| | Support (N) | 1486 | 1488 | 1478 | 1482 | 1490 | |
| Seresnet152 | Precision | 100% | 100% | 100% | 100% | 100% | 99.81% |
| | Recall | 100% | 100% | 100% | 100% | 100% | |
| | F1-score | 100% | 100% | 100% | 100% | 100% | |
| | Support (N) | 1484 | 1485 | 1482 | 1490 | 1483 | |
| Densenet201 | Precision | 99% | 99% | 99% | 100% | 100% | 99.45% |
| | Recall | 99% | 99% | 100% | 100% | 100% | |
| | F1-score | 99% | 99% | 99% | 100% | 100% | |
| | Support (N) | 1482 | 1484 | 1482 | 1490 | 1486 | |
| Xception | Precision | 98% | 97% | 98% | 100% | 99% | 98.31% |
| | Recall | 97% | 97% | 98% | 100% | 99% | |
| | F1-score | 98% | 97% | 98% | 100% | 99% | |
| | Support (N) | 1483 | 1481 | 1485 | 1487 | 1488 | |
| MobilenetV2 | Precision | 100% | 99% | 99% | 100% | 100% | 99.53% |
| | Recall | 99% | 99% | 99% | 100% | 100% | |
| | F1-score | 99% | 99% | 99% | 100% | 100% | |
| | Support (N) | 1483 | 1484 | 1489 | 1482 | 1486 | |

The performance of six transfer learning CNN architectures, VGG19, ResNet152V2, SE-ResNet152, DenseNet201, Xception, and MobileNetV2, is summarised in Table 4. DenseNet201 achieved the highest accuracy (99.45%), while VGG19 showed the lowest (88.58%). Transfer learning models generally demonstrated lower accuracy compared to their original CNN counterparts.

Table 4 presents the Precision, Recall, and F1-score results of CNN networks with transfer learning. DenseNet201 achieved the highest precision (99%), while SE-ResNet152 showed the best overall performance with 99.81% accuracy. VGG19 had the lowest precision (87%).

Dyskeratotic=(Dyk) Koilocytotic=(Koc) Metaplastic= (Mep) Parabasal=(Pab) Superficial Moderate=(Sfi).

**VGG19**

|     | Dyk  | Koc  | Mep  | Pab  | Sfi  |
|-----|------|------|------|------|------|
| Dyk | 1212 | 114  | 62   | 10   | 0    |
| Koc | 116  | 1190 | 57   | 3    | 41   |
| Mep | 113  | 101  | 1313 | 4    | 45   |
| Pab | 32   | 28   | 25   | 1471 | 3    |
| Sfi | 11   | 47   | 30   | 0    | 1396 |

**DenseNet201**

|     | Dyk  | Koc  | Mep  | Pab  | Sfi  |
|-----|------|------|------|------|------|
| Dyk | 1467 | 9    | 1    | 0    | 0    |
| Koc | 8    | 1467 | 3    | 0    | 0    |
| Mep | 7    | 5    | 1478 | 0    | 0    |
| Pab | 0    | 0    | 0    | 1490 | 0    |
| Sfi | 0    | 3    | 0    | 0    | 1486 |

**ResNet152v2**

|     | Dyk  | Koc  | Mep  | Pab  | Sfi  |
|-----|------|------|------|------|------|
| Dyk | 1476 | 7    | 4    | 0    | 2    |
| Koc | 5    | 1458 | 7    | 0    | 4    |
| Mep | 4    | 14   | 1462 | 0    | 1    |
| Pab | 0    | 1    | 1    | 1482 | 0    |
| Sfi | 1    | 8    | 4    | 0    | 1483 |

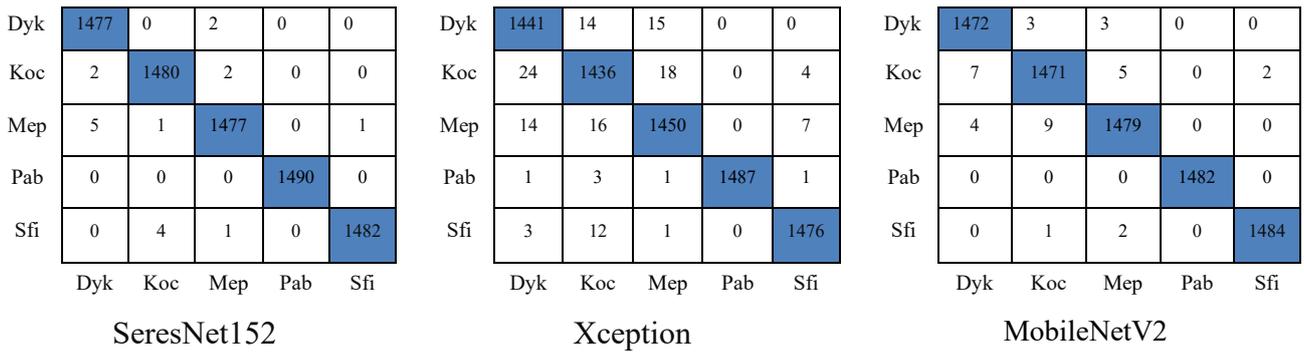

Figure 13: Six confusion matrices of Transfer learning CNNs.

Figure 13 shows the confusion matrices of transfer learning CNNs. DenseNet201 and SE-ResNet152 achieved the highest accuracy with minimal misclassification, while VGG19 had the most errors, especially in distinguishing Koc and Dyk cells.

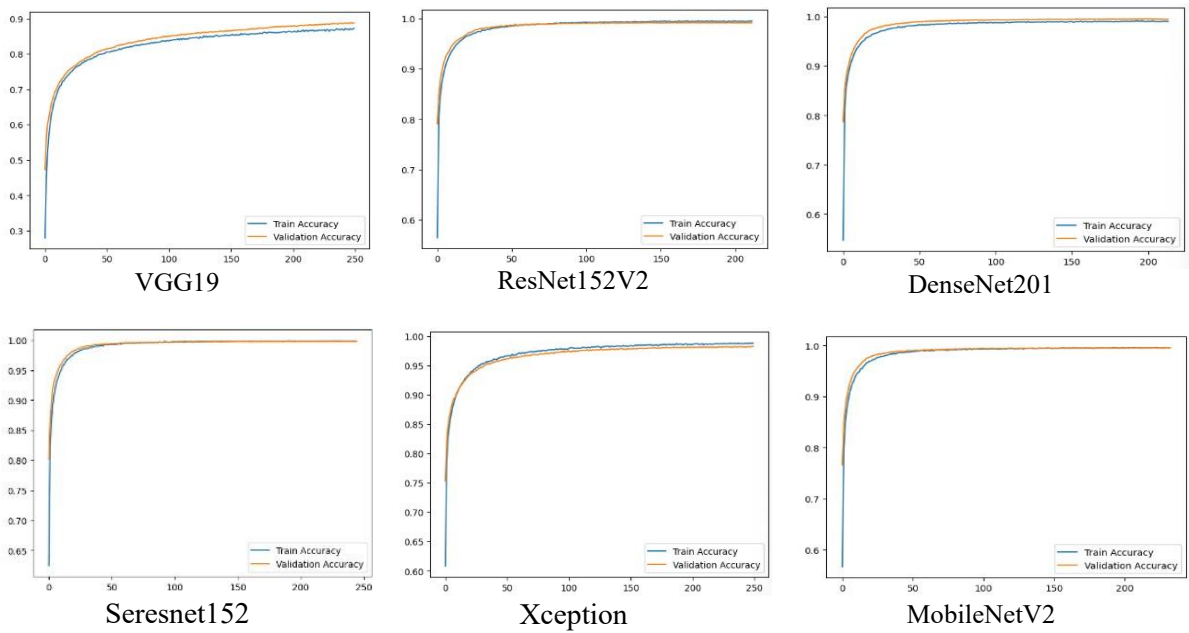

Figure 14: The Transfer Learning version's training and validation accuracy.

Figure 14 shows the training and validation accuracy of various Transfer Learning models. DenseNet201 and SE-ResNet152 demonstrate the highest accuracy with quick convergence and stability, while ResNet152V2 shows slightly lower accuracy compared to the others.

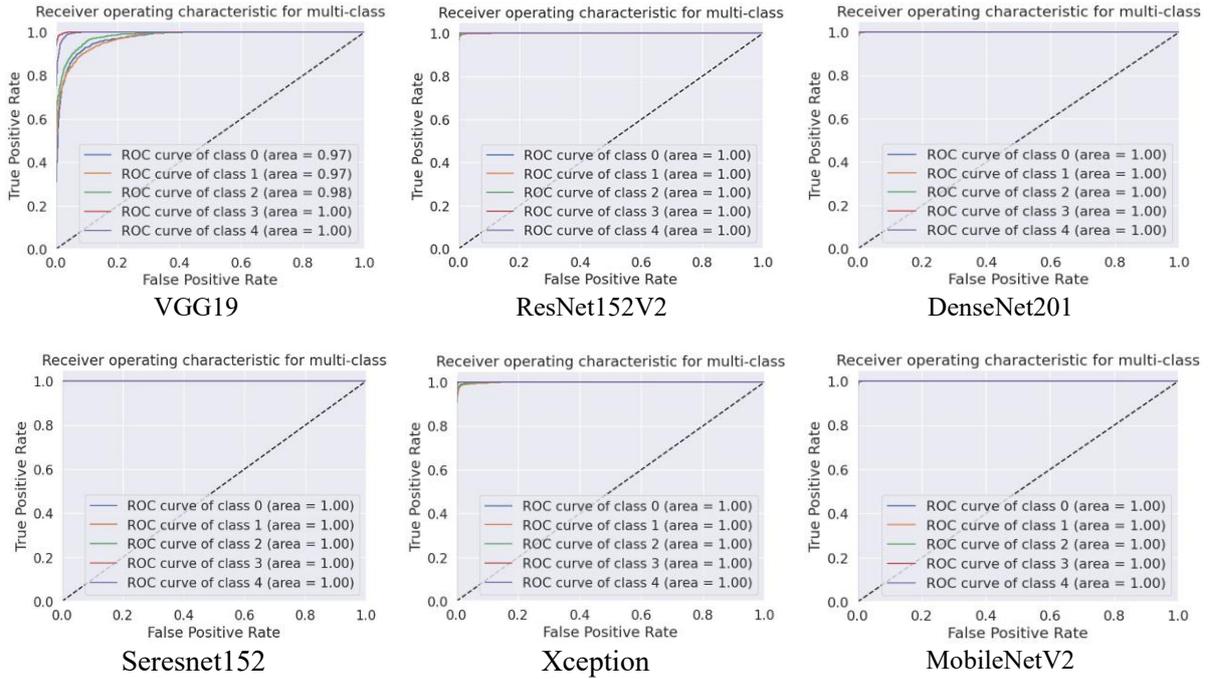

Figure 15: The Transfer Learning version's training and validation accuracy.

Figure 15 presents the ROC curves for various Transfer Learning models in multi-class classification. DenseNet201, SENet152, Xception, and MobileNetV2 show perfect classification performance with an AUC of 1.00 across all classes. VGG19, however, has a slightly lower AUC for one class.

## 4.4 Result of the XAI

This study uses three (3) explainable AI techniques, LIME, SHAP, and Grad-CAM, to generate local and global explanations for the S-NET (the developed CNN) predictions on the validation and test sets. LIME generates an interpretable model by training a local linear model around the prediction point, while SHAP provides a unified framework for feature importance estimation. Furthermore, we conducted statistical analysis on the correctly classified, misclassified and true positive-true negative-false positive-false negative images.

### 4.4.1 Visualization using LIME and SHAP

LIME and SHAP are used to interpret the S-Net cervical cancer model. LIME breaks the input images into super-pixel regions and approximates the model's behavior using a simple model like linear regression. The resulting heatmap highlights regions that positively or negatively impact the prediction, with red areas indicating strong influence and blue areas decreasing the likelihood of a class.

SHAP utilizes a pre-trained cervical cancer model to compute Shapley values, which assess the contribution of each pixel or region to the predicted class. By perturbing the input image and predicting the outputs for each variation, SHAP provides a quantitative explanation of the model's decisions.

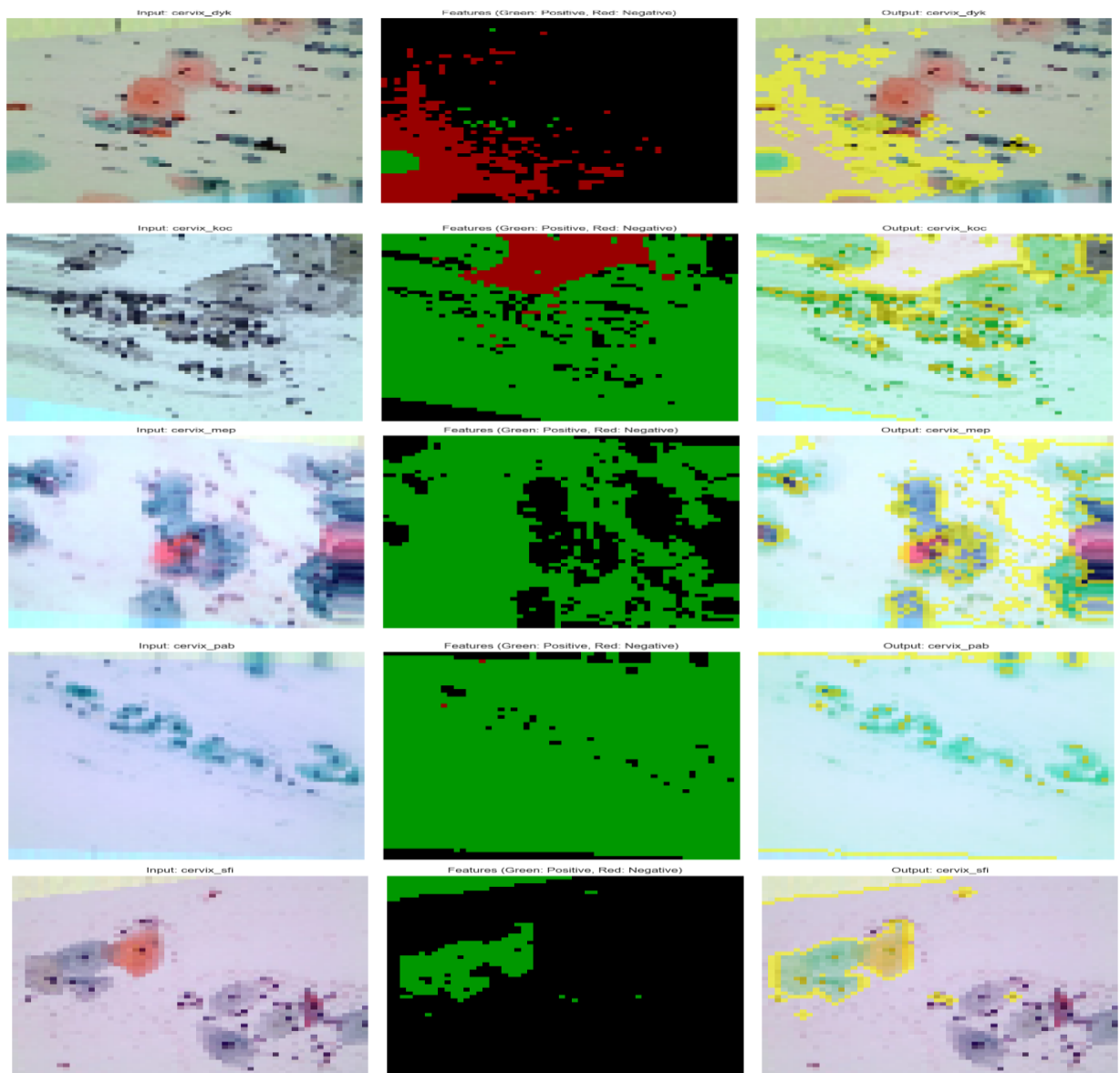

Figure 16: LIME partition explainer of Pap smear images

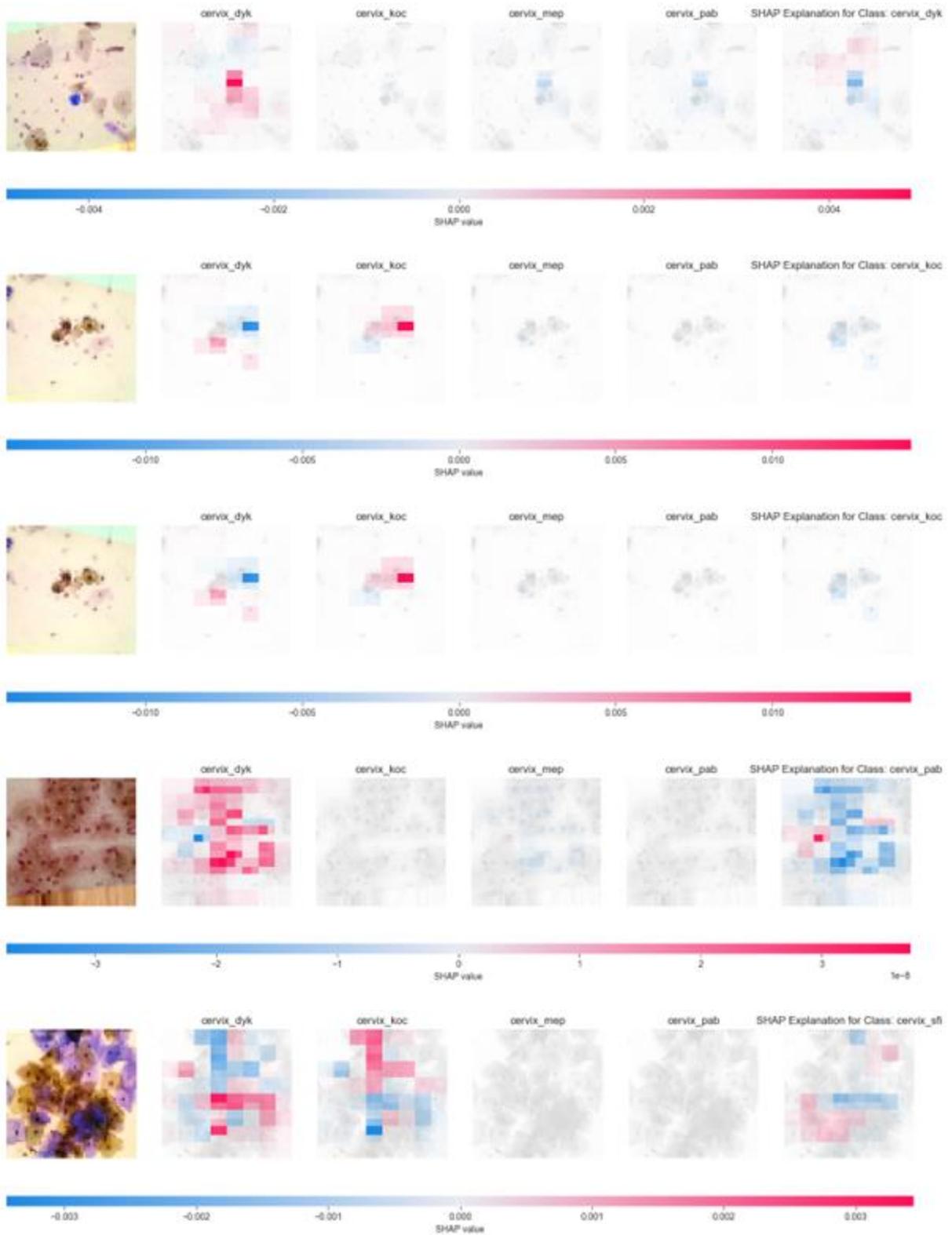

Figure 17: SHAP explainer of Pap Smear images

## 4.4.2 Grad-CAM analysis of correctly/incorrectly classified cervical cancer modalities

Grad-CAM is a technique that visualizes which regions of an image contribute most to the S-Net model's prediction by generating heatmaps. It computes the gradients of the predicted class concerning the feature maps in the model's last convolutional layer. These gradients highlight the importance of each feature map, which are then weighted and combined to create a class activation heatmap. The heatmap is overlaid on the original image, with red regions indicating higher attention and blue regions indicating less influence on the prediction. This technique enhances the transparency and interpretability of the S-Net model in cervical cancer classification, allowing for a better understanding of how the model focuses on specific areas. Misclassification examples are displayed in Figure 17.

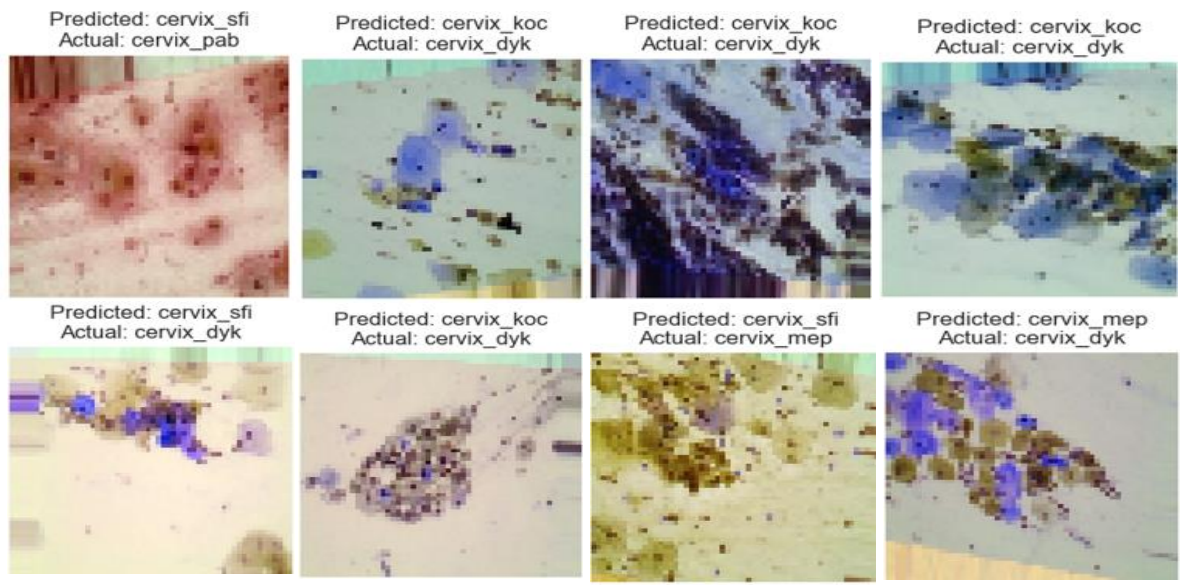

Figure 18: Examples of misclassified images

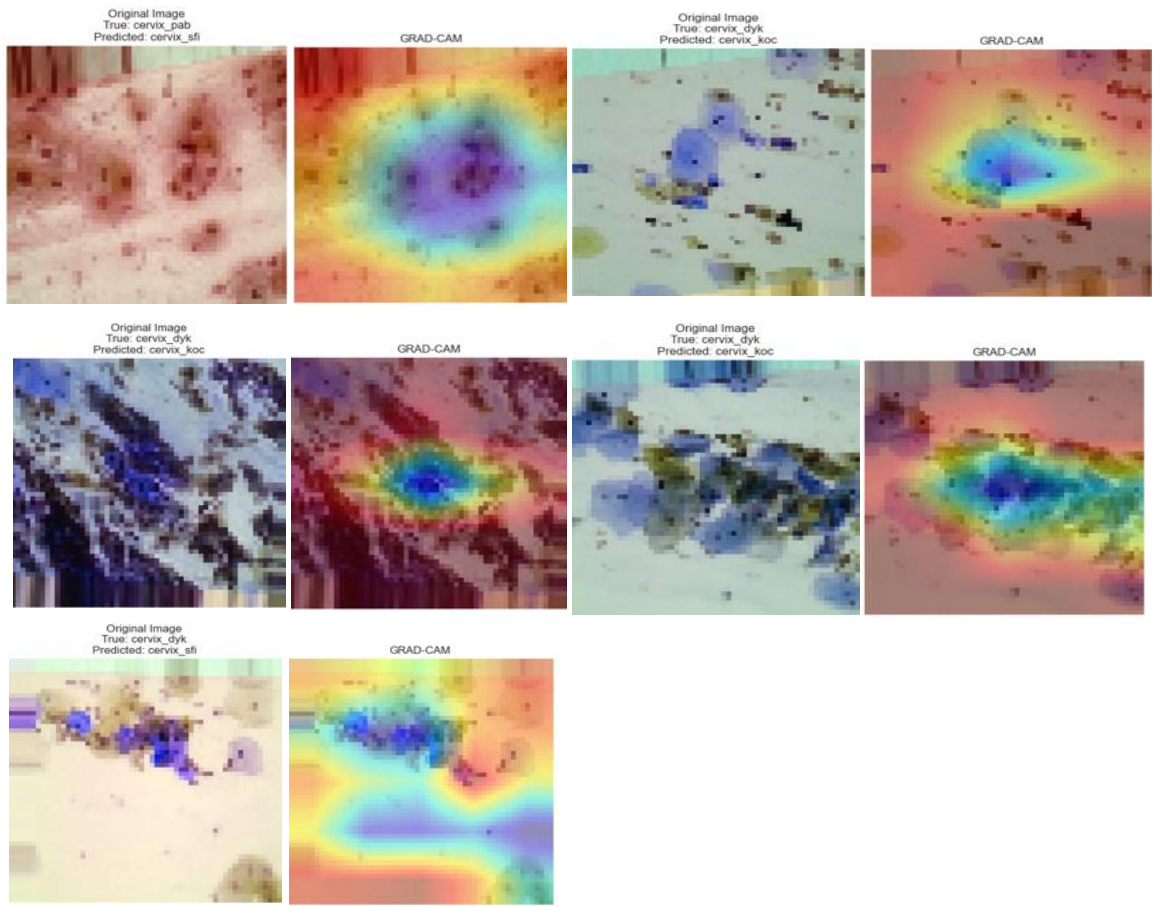

Figure 19: GRAD-CAM view of misclassified images

Figure 19 presents Grad-CAM visualizations revealing misclassified cervical cancer cases. Red regions indicate high model attention, while blue regions show less focus. Misclassifications occurred when S-Net focused on irrelevant areas, such as predicting "cervix_Sfi" instead of "cervix_Koc." This suggests challenges in identifying cancer-specific features accurately.

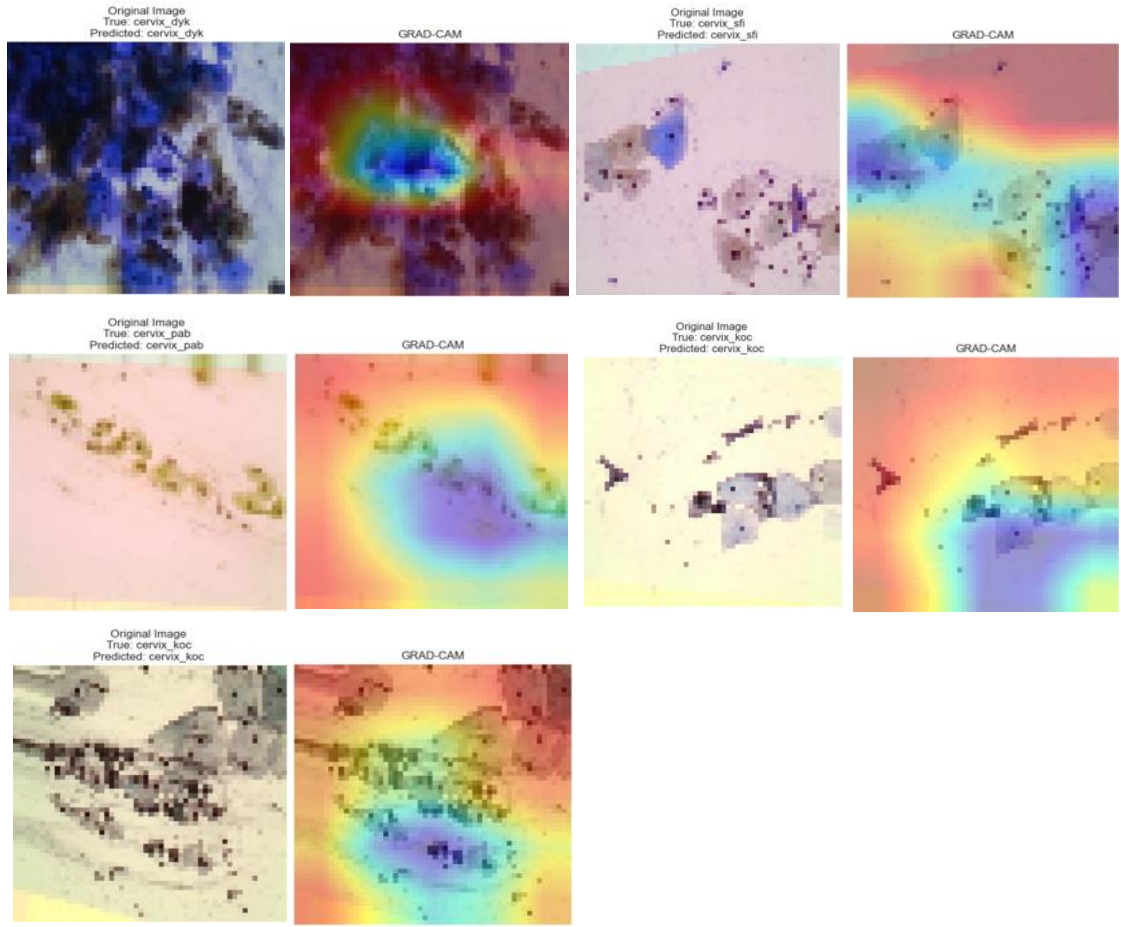

Figure 20: Grad-CAM view of correctly classified images

Figure 20 presents Grad-CAM visualizations of correctly classified cervical cell images. The highlighted regions (in red) indicate where the model focuses during classification, aligning well with class-specific features. This confirms the model's ability to extract and utilize meaningful visual cues. However, occasional focus on irrelevant regions (in blue) suggests opportunities to further refine spatial attention and enhance precision.

### 4.4.3 Pixel intensity of correctly/incorrectly Cervical Cancer modalities

Pixel intensity highlights areas that contribute most to the CNN's decision. The original image shows key features, while the Gradient × Input explanation reveals the influence of each part, with brighter areas indicating higher influence. Misalignment between the highlighted regions and expected features suggests the model may focus on irrelevant areas, impacting accurate classification.

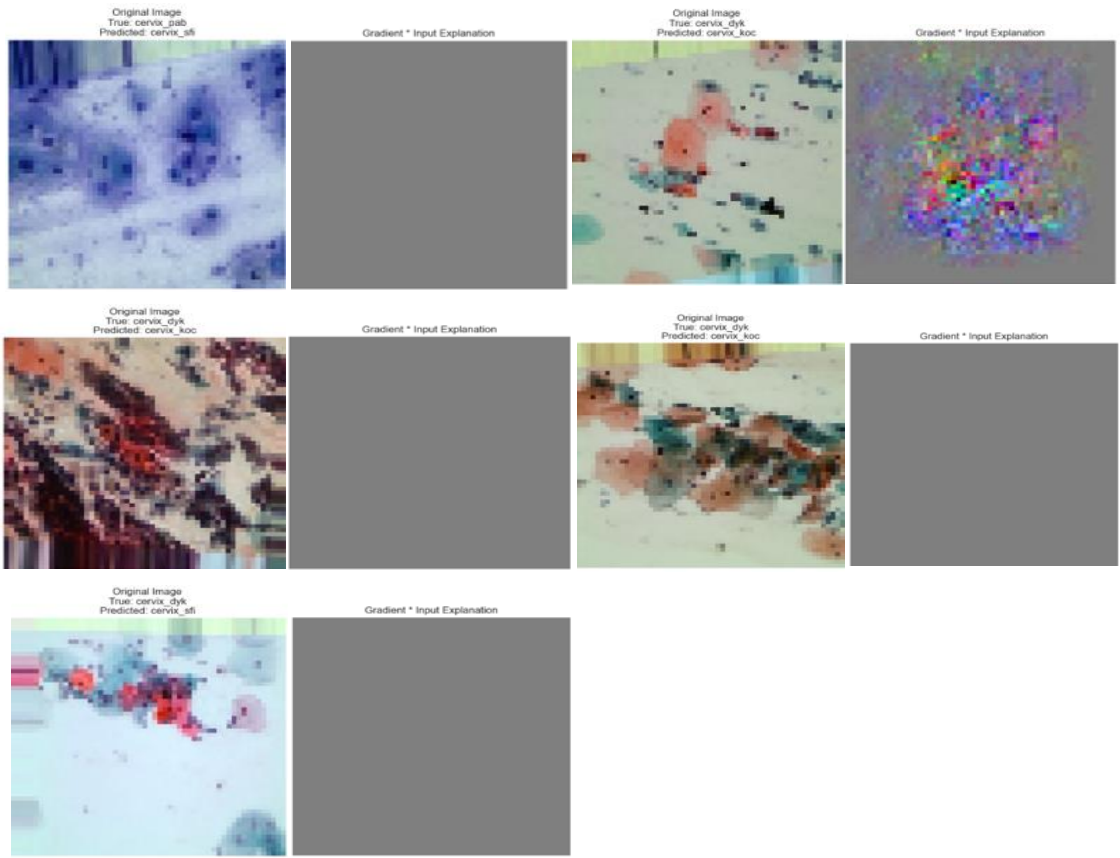

Figure 21: Grad-CAM view of pixel intensity of misclassified images

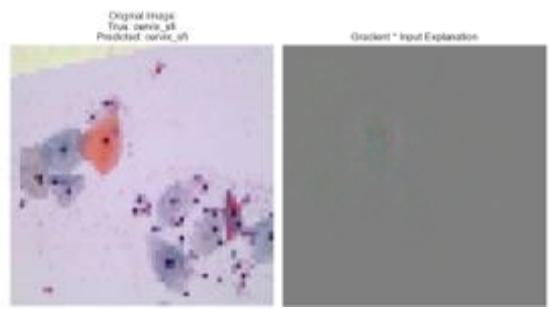

Figure 22: Grad-CAM view of pixel intensity for correctly classified images

### 4.4.4  Pixel Intensity Analysis Correct/misclassified images

This study compares the pixel intensity distributions between correctly classified and misclassified cervical cancer images. The analysis reveals that correctly classified images have significantly higher mean and median pixel intensities (mean = 61.02, median = 67.24) compared to misclassified images (mean = 44.94, median = 49.33). The interquartile range (IQR) is also larger for the correctly classified images, indicating more variability in their pixel intensities. In contrast, misclassified images exhibit lower variability, suggesting that the model struggles to classify images with less distinct or variable features. This emphasizes the crucial role of clear and distinct pixel intensity contrasts for accurate cervical cancer detection, as the model relies on such intensity patterns to make accurate predictions.

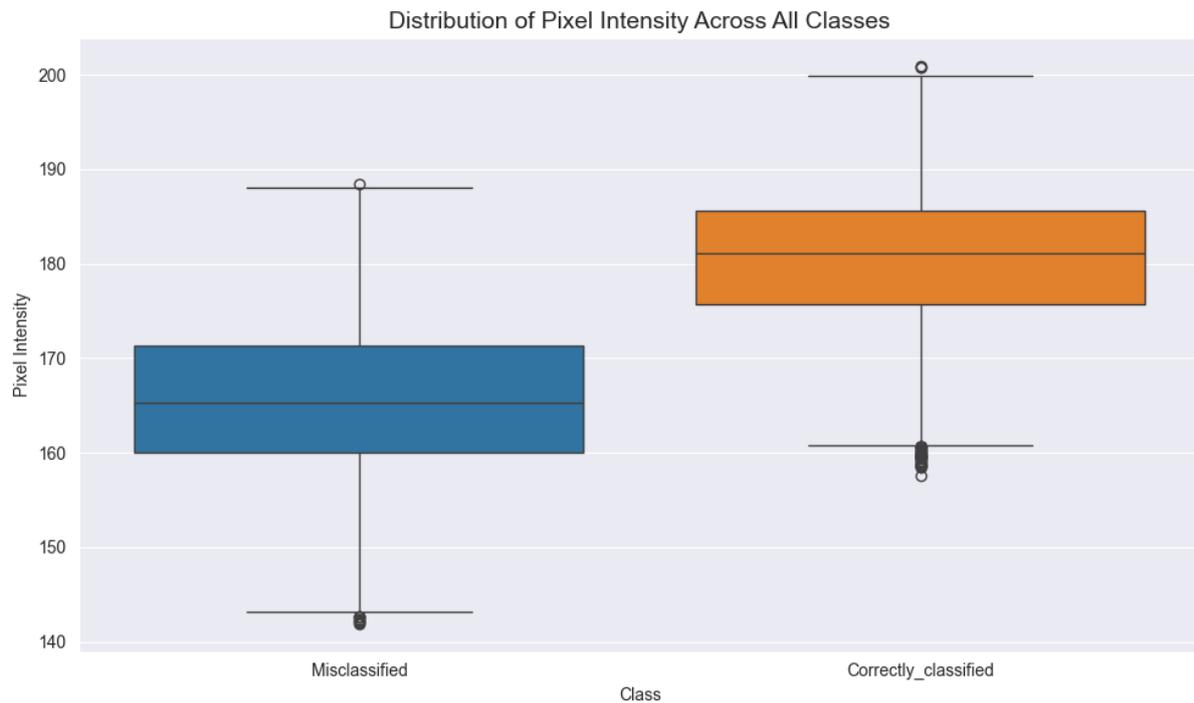

Figure 23: Distribution of Pixel intensity of misclassified/ correctly classified images.

Furthermore, we use Significance Level (α) = 0.05 as the threshold to determine statistical significance. If the p-value is less than α, we reject the null hypothesis and conclude a significant difference between the two classes. If the p-value exceeds α, we fail to reject the null hypothesis, indicating no significant difference.

**Result:**

Independent t-test p-value: 0.0000

Mann–Whitney U test p-value: 0.0000

Independent t-test: Reject H₀. There is a significant difference between misclassified_100 and correctly_classified_100.

Table 5: Descriptive Statistics for Pixel Intensity of correctly classified and misclassified images.

| Statistic | Misclassified | Correctly Classified |
|---|---|---|
| **Count** | 4096.000000 | 4096.000000 |
| **Mean** | 44.938873 | 61.021790 |
| **Std** | 29.812426 | 36.996021 |
| **Min** | 0.416000 | 4.292000 |
| **25%** | 12.529000 | 22.864000 |
| **50%** | 49.338001 | 67.242001 |
| **75%** | 75.280998 | 98.147999 |
| **Max** | 85.456001 | 112.676003 |
| **Interquartile Range** | 62.751998 | 75.283998 |

### 4.4.5 Pixel intensity of TP-FP-TN-FN Cervical Cancer modalities

The analysis of pixel intensity distributions across four categories (True Positive, True Negative, False Positive, and False Negative) reveals key insights. False Negatives (FN) have the highest mean intensity (62.04) and the broadest range, indicating errors arise from highly variable intensities. True Positives (TP) and True Negatives (TN) show more consistent distributions, suggesting better performance when intensities align with patterns. False Positives (FP) exhibit a narrower intensity range, with errors occurring in areas with less contrast. Overall, the model performs best with mid-range intensities and struggles with extreme or variable intensities, especially in the FN category.

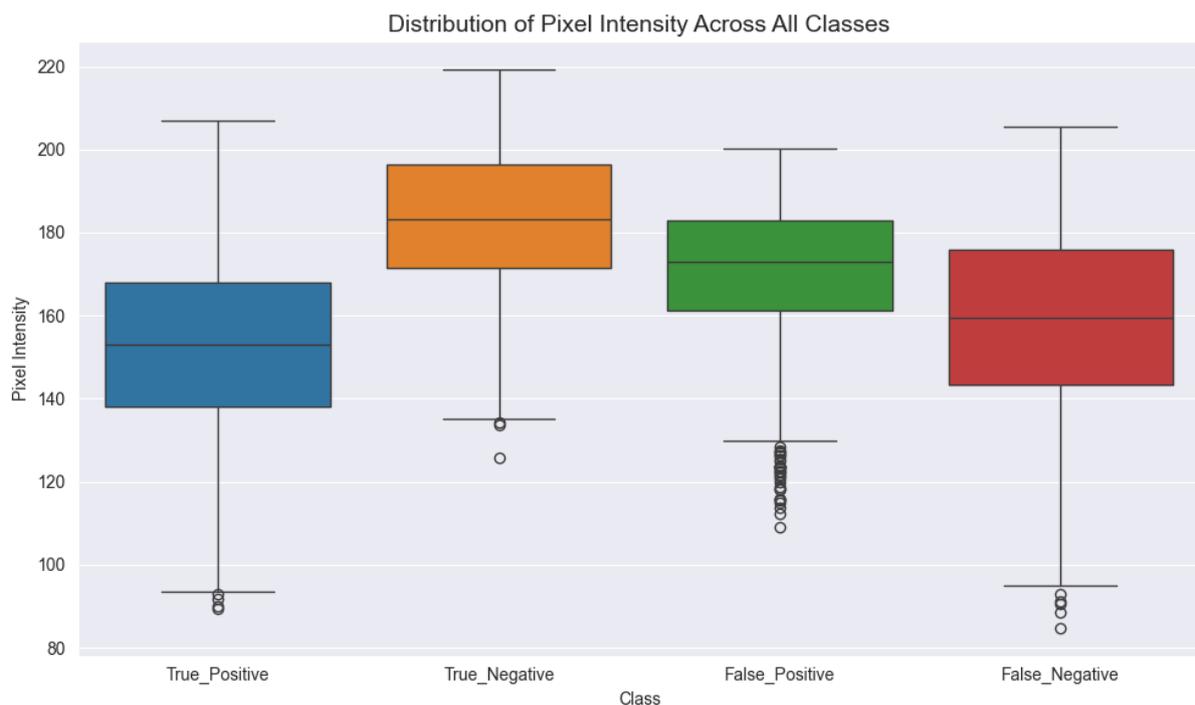

Figure 24: Distribution of Pixel intensity of TP-TN-FP-FN

## 4.4.6 Pixel intercity comparison between TP-FP-TN-FN

Significance Level (α): We use α = 0.05 as the threshold to determine statistical significance. If the p-value is less than α, we reject the null hypothesis and conclude that there is a significant difference between the two classes. If the p-value exceeds α, we fail to reject the null hypothesis, indicating no significant difference.

**Accepted/rejected Hypotheses:**

Independent t-test and Mann–Whitney U test: Reject $H1_o$: The pixel intensity distributions of True positive and True negative are not significantly different.

Independent t-test and Mann–Whitney U test: Reject $H2_o$: The pixel intensity distributions of True positive and False positive are not significantly different.

Independent t-test and Mann–Whitney U test: Reject $H3_o$: The pixel intensity distributions of True positive and False negative are not significantly different.

Independent t-test and Mann–Whitney U test: Reject $H4_o$: The pixel intensity distributions of True negative and False positive are not significantly different.

Independent t-test and Mann–Whitney U test: Reject $H5_o$: The pixel intensity distributions of True negative and False negative are not significantly different.

Independent t-test and Mann–Whitney U test: Reject $H6_o$: The pixel intensity distributions of False positive and False negative are not significantly different.

**Descriptive Statistics for Pixel Intensity:**

Table 6: Descriptive Statistics for Pixel Intensity of TP-TN-FP-FN images

| Statistic | True Positive | True Negative | False Positive | False Negative |
|---|---|---|---|---|
| **Count** | 3136.000000 | 3136.000000 | 3136.000000 | 3136.000000 |
| **Mean** | 152.820465 | 183.763855 | 170.796997 | 158.757401 |
| **Std** | 21.152983 | 17.148066 | 15.314829 | 22.917040 |
| **Min** | 89.400002 | 125.800003 | 109.000000 | 84.800003 |
| **25%** | 138.199997 | 171.600006 | 161.199997 | 143.199997 |
| **50%** | 153.000000 | 183.299995 | 173.000000 | 159.600006 |
| **75%** | 168.000000 | 196.399994 | 182.850002 | 175.800003 |
| **Max** | 207.000000 | 219.199997 | 200.199997 | 205.399994 |
| **Interquartile Range** | 29.800003 | 24.799988 | 21.650005 | 32.600006 |

## 5. Discussion

In this study, a lightweight convolutional neural network (CNN) model, termed S-Net, was developed with fewer layers and learnable parameters to efficiently detect cervical cancer in Pap smear images. The results demonstrate that the proposed S-Net model effectively classifies cervical cancer cells and achieves an impressive accuracy of 99.99%, outperforming state-of-the-art (SOTA) CNN models, transfer learning techniques, and ensemble methods. For training and evaluation, the S-Net model utilized three Python data generation techniques: TensorFlow/Keras's flow_from_directory, flow_from_dataframe, and flow (used with NumPy arrays). Among these methods, flow_from_directory yielded the best performance, as it efficiently loads images batch-by-batch, minimizing memory consumption compared to flow_from_dataframe and flow, which load the entire dataset into RAM. Furthermore, the Adam optimizer provided consistent and stable performance across all experiments, while the RMSprop optimizer showed fluctuations depending on the model, highlighting the significance of selecting the appropriate optimizer based on the specific convergence behavior of the architecture.

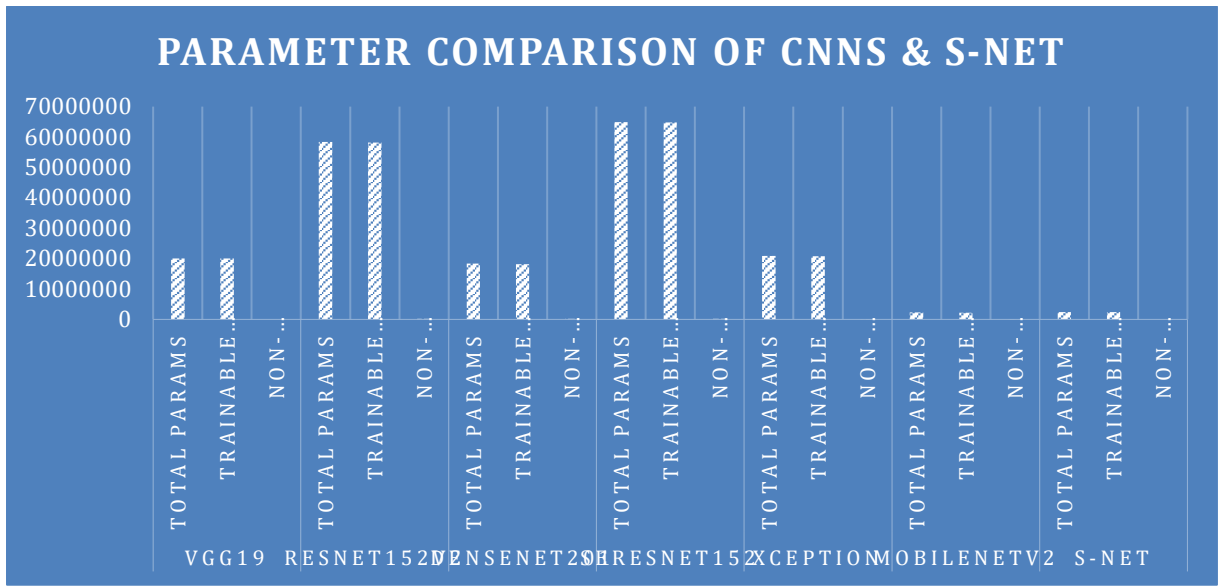

Figure 25: Accuracy comparison among individual CNN, transfer learning, and Mnet models

This research presents three (3) experiments for detecting and classifying cervical cancer images using well-known deep learning architectures and identifying the most promising model. In the first experiment, six CNN models, VGG19, ResNet152V2, SE-ResNet152, DenseNet201, Xception, and MobileNetV2, were compared on 25,000 microscopic cervical cancer images across five (5) classes: Dyk, Koc, Mep, Pab, and Sfi.

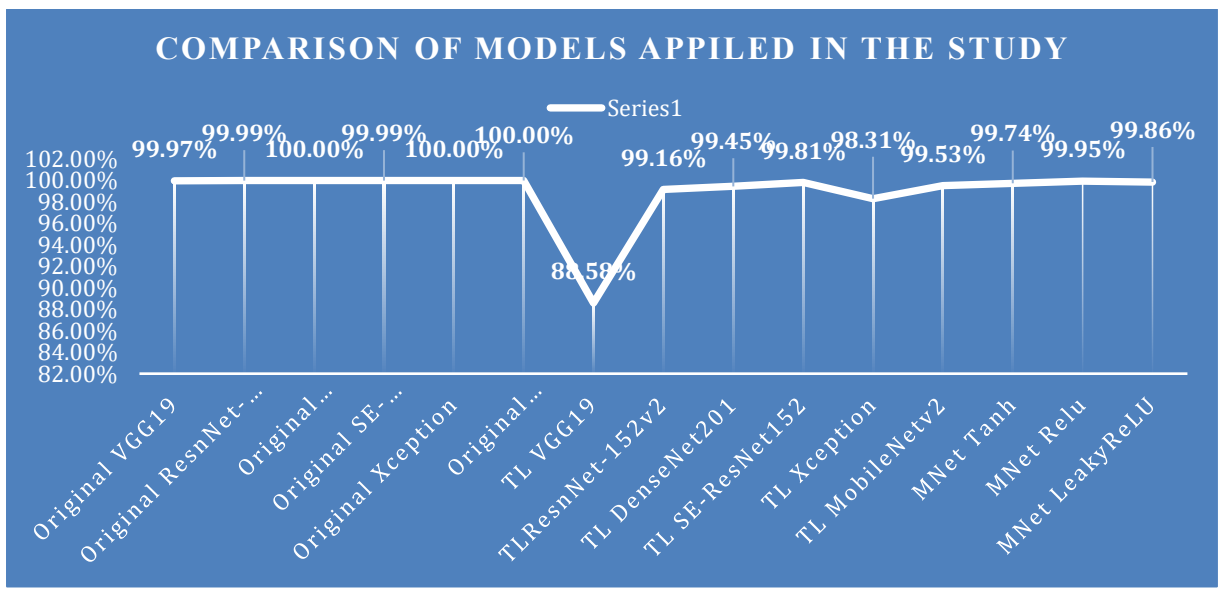

Figure 26: Accuracy comparison among individual CNN, transfer learning, and ensemble models.

Significant studies have highlighted that transfer learning improves classification accuracy and reduces training time compared to conventional CNN for many classification tasks (Ju et al., 2022; Karimi et al., 2021). For example, Mehmood et al. (2024) proposed and reported 95.07% binary classification performance; Emam et al. (2024) reported breast cancer diagnosis using an optimized transfer learning technique and improved the accuracy. However, in our study, transfer learning received negative results. The accuracy has decreased compared to the main CNN. The findings of negative transfer learning support the study (Rangarajan & Raja, 2020; Mohanty et al., 2016; Barbedo, 2018) that in the case the input image differs from the trained data of the Imagenet Dataset, the accuracy is likely to be decreased. The effect of background noise and the application of different augmentation techniques separately with the test sets resulted in a drop in performance. However, in the case of the original CNN, the model was trained and tested using similar input, and the prediction capabilities were increased in unseen data. Moreover, although CNN can learn features irrespective of the input data, this study's limited number of datasets is likely a factor influencing the prediction capability. Our view is also supported by (Barbedo, 2018), who suggested that increasing the dataset size may improve transfer learning performance when the input image is modified using augmentation.

Most importantly, this study applied three (3) XAI techniques, LIME, SHAP, and Grad-CAM, to generate explanations for CNN detection and classification. The XAI expands our understanding of S-NET's decision-making process. XAIs used in this study enhance the interpretability of S-Net. LIME provides model-agnostic explanations by approximating the local decision boundary of a model with an interpretable surrogate, highlighting the regions of the pap smear images that contribute most to the model's predictions. Based on cooperative game theory, SHAP offers consistent and theoretically sound explanations by assigning Shapley values to input features, thereby quantifying their contribution to the output with high fidelity. Grad-CAM, designed explicitly for CNNs, generates class-discriminative visualizations by leveraging gradients from the final convolutional layers, effectively localizing regions in the image that influence a specific class prediction. In summary, XAI, coupled with S-Net, enhances the trust and understanding of CNNs by transparentizing their decision-making processes.

Lastly, pixel intensity was analyzed using the statistical method of true classification, misclassification, true positive, false positive, true negative, and false negative classification. This study contributes to understanding the growing literature on using explainable AI to improve medical image analysis and diagnosis. It provides insights into the interpretability and transparency of CNN for Cervical cancer modalities.

## 6. Conclusion and future work

This study evaluates deep learning models for cervical cancer detection, conducting three sequential experiments. The first experiment tested six CNN architectures—VGG19, ResNet152V2, DenseNet201, ResNeXt101, SE-ResNet152, and MobileNetV2—on a dataset of 25,000 Pap smear images. The second experiment applied transfer learning to these models. The third introduced the novel lightweight CNN, S-Net, and integrated it with Explainable AI (XAI) methods: LIME, SHAP, and Grad-CAM.

The results demonstrate that S-Net outperformed all baseline models, achieving superior accuracy, precision, recall, and F1-score. XAI techniques improved interpretability, highlighting critical regions influencing predictions. A pixel intensity analysis revealed significant differences between correctly and incorrectly classified samples, emphasizing the role of intensity in model performance.

However, the study acknowledged limitations, including the use of a secondary dataset and the exclusive focus on Pap smear images, limiting generalizability. Future work should explore other imaging modalities like colposcopy and histopathology to enhance clinical applicability. While transfer learning did not yield optimal results, further research on lightweight CNNs may be beneficial. Clinical validation with expert input is essential for real-world deployment.

In conclusion, this research underscores the potential of CNNs, particularly S-Net, in automating cervical cancer detection, offering a significant contribution toward reliable and interpretable AI-based medical diagnostics